\documentclass[aps,prb,reprint,showpacs,superscriptaddress]{revtex4-2}
\usepackage{graphicx}
\usepackage{dcolumn}
\usepackage{bm}
\usepackage{amssymb}
\usepackage{amsmath}
\usepackage{subfigure}
\usepackage{epstopdf}
\usepackage{float}
\usepackage{mathrsfs}
\usepackage{xcolor}
\usepackage{yhmath}
\usepackage{mathrsfs}

\begin{document}
\title{Exceptional spin wave dynamics in an antiferromagnetic honeycomb lattice}

\author{Yong Wang}
\email[]{yongwang@nankai.edu.cn}
\affiliation{School of Physics, Nankai University,  Tianjin 300071, China}

\author{Wang Yao}
\affiliation{Department of Physics and Center of Theoretical and Computational Physics,  The University of Hong Kong,  Hong Kong,  China}

\begin{abstract} 
We theoretically investigate possible effects of electric current on the spin wave dynamics for the N\'{e}el-type antiferromagnetic order in a honeycomb lattice.  Based on a general vector decomposition of the spin polarization of conduction electrons,  we find that there can exist reciprocal and nonreciprocal terms in the current-induced torque acting on the local spins in the system.   Furthermore,  we show that the reciprocal terms will cause the spin wave Doppler effect,  while the nonreciprocal terms can induce rich non-Hermitian topological phenomena in the spin wave dynamics,  including exceptional points,  bulk Fermi arc,  non-Hermitian skin effect,  etc.  Our results indicate the capability to manipulate non-Hermitian magnons in magnetic materials by electric current,  which could be important for both fundamental physics and technology applications.
\end{abstract}
\maketitle
\emph{Introduction}  
Spin wave,  whose quanta is termed ``magnon",  describes the collective motion of magnetic moments in magnetic materials\cite{Callaway}.  The dispersion of spin wave modes,  or magnon spectrum alternatively,  depends on the effective interactions between the magnetic moments.   Therefore,  measuring the spin wave dynamics can provide indispensable information about the magnetic orders in materials.  Moreover,  the spin waves can be excited and tuned by different approaches.  For example,  the injection of an electric current into the magnetic materials can cause the coherent excitation\cite{JMMM1999,PRL1998,Nature2000, PRL2003,PRL2012Xiao,NC2018Akerman},   Doppler shift\cite{PRB1998,MacDonald2004,Science2008,PRB2010,PRL2012},  attenuation suppresion and amplification\cite{PRL2009,PRL2014} of the spin waves.   In fact,  the capability to control spin waves by the electric current is essential for the implementation of wave-based computing\cite{JAP2020} and the development of magnon-based spintronics technologies\cite{NC2014,NatPhys2015}.  

The recent breakthrough to fabricate two-dimensional (2D) magnetic materials\cite{Nature1,Nature2} has triggered intensive activities to explore the spin wave dynamics in the 2D limit.  For example,  the spin wave modes in the 2D ferromagnetic material CrI$_{3}$ have been observed by micro-Raman spectroscopy\cite{NC2018} and inelastic neutron scattering technique\cite{PRX2018},  respectively.   In the van der Waals antiferromagnetic(AFM) materials,   the magnon transport over several micrometers has been experimentally observed\cite{PRX2019},  and the spin Nernst effect due to the non-trival topology of magnon band has been theoretically discussed\cite{Nernst1,Nernst2}.  Furthermore,  the ability to tune spin wave by gate has been demonstrated in the AFM bilayer CI$_{3}$\cite{NatMat2020} and the magnon valve structure made of MnPS$_{3}$\cite{NC2021},  while the topological magnon insulator has been realized in CrXTe$_{3}$ (X = Si, Ge) compounds\cite{SA2021}.  

2D magnets,  in particular the AFM,  provide a new realm to explore the influence of electric current on the spin wave dynamics.  In conventional spin valve structures and ferromagnetic spin textures\cite{JMMM1999,PRL1998,Nature2000,PRL2003,PRL2012Xiao,NC2018Akerman,PRB1998,MacDonald2004,Science2008,PRB2010,PRL2012,PRL2009,PRL2014},  the magnetization configuration is slowly varying in space,  allowing it to be treated as a continuous vector field in formulating its dynamical control by the electric current via the spin torque effects\cite{STT1,  STT2, STTrev1,STTrev2,  PRB1998,ZhangLi2004}.  However,  such a methodology may not be appropriate to describe the current-induced spin dynamics when the direction of magnetic moments can change abruptly in the atomic scale.  In fact,  the longstanding disputes on the form of spin torque in the AFM materials \cite{AFMtorq1,AFMtorq2,AFMtorq3,AFMtorq4,AFMtorq5,AFMtorq6,AFMtorq7} already manifest the challenge to treat the antiparallel alignment of atomic magnetic moments between the sublattices.  Moreover,  the lack of atomic resolution in the long wavelength approximation implies that important phenomena of the spin wave spectrum in the whole Brillouin zone can be missed in this treatment.     

In this paper,  we exploit a vector analysis approach to study the spin wave dynamics of a 2D AFM lattice in presence of an electric current.   We show that the spin torque acting on each localized spin can be decomposed into reciprocal and nonreciprocal terms,  which can induce Doppler shift,  exceptional point,  bulk Fermi arc,  as well as non-Hermitian skin effect in the spin wave dynamics in this exemplary system.  Our results reveal the wide existence of current-induced non-Hermitian topological phenomena in magnetic materials,  which however have been overlooked by the conventional approaches for spin valve structures and magnetization textures. 

\begin{figure}[!ht]
  \centering
  \includegraphics[width=0.5\textwidth,clip]{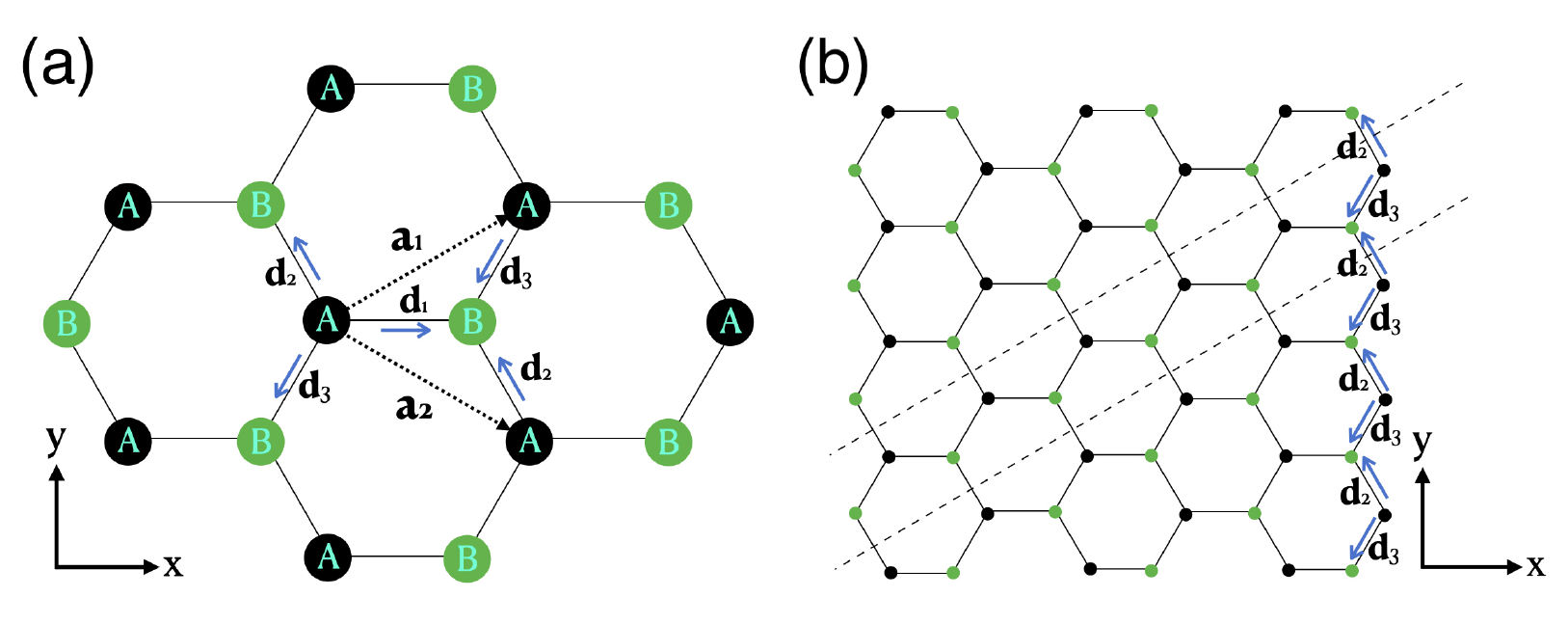}
  \caption{(Color online) (a) Crystal structure of an N\'{e}el-type AFM honeycomb lattice.  $\mathbf{a}_{1}$ and $\mathbf{a}_{2}$ denote the lattice vectors,  and $A$ and $B$ denote the two spin sublattices.  $\mathbf{d}_{1},\mathbf{d}_{2},\mathbf{d}_{3}$ denote the vectors from a $A$ site to its three nearest neighbouring  $B$ sites; (b) crystal structure of an AFM nanoribbon with zigzag edges.  One unit cell of the nanoribbon is enclosed by two dashed lines.} 
 \label{Fig1}
\end{figure}

\emph{Model Hamiltonian and Spin Dynamics}
The model system under consideration is an N\'{e}el-type AFM honeycomb lattice in presence of an in-plane electric current $\mathbf{I}_{e}$,  as shown schematically in Fig.~\ref{Fig1}(a).  The two lattice vectors are denoted as $\mathbf{a}_{1}=a(\frac{\sqrt{3}}{2},\frac{1}{2})$ and $\mathbf{a}_{2}=a(\frac{\sqrt{3}}{2},-\frac{1}{2})$ with the lattice constant $a$,  and there are two inequivalent sites $(A, B)$ in each unit cell; $\mathbf{d}_{1,2,3}$ are the three vectors from one $A$ site to its neighbouring $B$ sites.  Under the nearest neighbor approximation,  the Hamiltonian of the system is written as $H=H_{S}+H_{c}+V_{sd}$,  where\cite{Mahan}
\begin{eqnarray}
H_{S}&=&J\sum_{\langle i,j\rangle}\hat{\mathbf{S}}_{i}\cdot\hat{\mathbf{S}}_{j}-K\sum_{i}(\hat{S}_{i}^{z})^{2},\label{HamS}\\
H_{c}&=&t\sum_{\langle i,j\rangle,s}(c_{i,s}^{\dag}c_{j,s}+c_{j,s}^{\dag}c_{i,s}),\label{HamC}\\
V_{sd}&=&J_{sd}\sum_{\langle i\rangle,ss'}\hat{\mathbf{S}}_{i}\cdot c_{i,s}^{\dag}\hat{\bm{\sigma}}_{ss'}c_{i,s'}.\label{Vsd}
\end{eqnarray}
Here,  $\hat{\mathbf{S}}_{i}$ is the localized spin operator of the magnetic atom at site $i$,   $c_{i,s}^{\dag}(c_{i,s})$ is the creation (annihilation) operator of conduction electrons at site $i$ with spin $s$,  and $\hat{\bm{\sigma}}$ are the Pauli matrices;  $J>0$ is the Heisenberg exchange coefficient between the localized spins located at two nearest neighbor sites $\langle i,j\rangle$,  and $K>0$ is the uniaxial magnetic anisotropy coefficient of localized spins,  $t$ is the hoping parameter of conduction electrons between the sites $\langle i,j\rangle$,  and $J_{sd}$ is the exchange coefficient between localized spins and conduction electrons.  

With the Hamiltonian $H$ above,  the classical dynamics of localized spin at site $i$ will be\cite{SMater} $\frac{d}{dt}\mathbf{S}_{i}=\frac{1}{\hbar}\bm{\mathcal{B}}_{i}\times\mathbf{S}_{i}$,  where $\hbar$ is the reduced Planck constant,  and the effective field acting on $\mathbf{S}_{i}$ takes the form $\bm{\mathcal{B}}_{i}=J\sum\limits_{\langle j\rangle_{i}}\mathbf{S}_{j}-2KS_{i}^{z}\hat{\mathbf{e}}_{z}+J_{sd}\bm{\sigma}_{i}$.   Here,  $\langle j\rangle_{i}$ denote the nearest neighbor site $j$ of the site $i$,  and $\bm{\sigma}_{i}=(\langle c_{i,\uparrow}^{\dag}c_{i,\downarrow}\rangle,\langle c_{i,\downarrow}^{\dag}c_{i,\uparrow}\rangle,\langle c_{i,\uparrow}^{\dag}c_{i,\uparrow}\rangle-\langle c_{i,\downarrow}^{\dag}c_{i,\downarrow}\rangle)$ is the spin polarization of conduction electrons at site $i$.   In the equilibrium case,  $\bm{\sigma}_{i}$ will vanish or align to the direction of $\mathbf{S}_{i}$,  which will not affect the dynamics of $\mathbf{S}_{i}$.  However,  when a non-equilibrium distribution of conduction electrons is established after applying the in-plane electric current,  $\bm{\sigma}_{i}$ can deviate from the direction of $\mathbf{S}_{i}$,  which will induce a spin torque on $\mathbf{S}_{i}$ to modify its dynamics.  

The non-equilibrium spin polarization $\bm{\sigma}_{i}$ depends on the local configurations of spins,  quantum transport processes of conduction electrons,  as well as the dissipation effects of surrounding environments,  which vary from materials to materials.  Instead of a material specific discussion,  here we exploit a vector analysis method to reveal the general features of the current-induced spin dynamics\cite{SMater}.  The non-equilibrium part of conduction electron's spin polarization at site $i$ is due to the microscopic current flows between $i$ and its neighbors $\langle j\rangle_{i}$,  on which the configurations of local spins are imprinted.  Therefore,  it is reasonable to assume that $\bm{\sigma}_{i}=\sum\limits_{\langle j\rangle_{i}}\bm{\sigma}_{i\leftarrow j}$,  where the contribution $\bm{\sigma}_{i\leftarrow j}$ depends on the magnetic moment of the site $\langle j\rangle_{i}$\cite{SMater}.  Furthermore,  we use the fact that each vector $\bm{\sigma}_{i\leftarrow j}$ can be decomposed in a Cartesian coordinates as $\bm{\sigma}_{i\leftarrow j}=\alpha_{i\leftarrow j}S^{-1}\mathbf{S}_{i}+\beta_{i\leftarrow j}S^{-2}\mathbf{S}_{i}\times\mathbf{S}_{j}+\gamma_{i\leftarrow j}S^{-3}\mathbf{S}_{i}\times(\mathbf{S}_{i}\times\mathbf{S}_{j})$.  Here,  $S$ is the magnitude of each localized spin.  After some vector algebra operations,  the spin torque on $\mathbf{S}_{i}$ due to $\bm{\sigma}_{i\leftarrow j}$ will be\cite{SMater}
\begin{eqnarray}
\bm{\tau}_{i\leftarrow j}=\frac{\beta_{i\leftarrow j}J_{sd}}{\hbar S^{2}}(\mathbf{S}_{i}\times\mathbf{S}_{j})\times\mathbf{S}_{i}+\frac{\gamma_{i\leftarrow j}J_{sd}}{\hbar S}\mathbf{S}_{i}\times\mathbf{S}_{j}.\label{tauij}
\end{eqnarray} 
Note that Eq.~(\ref{tauij}) resembles the forms of damping-like and field-like torque in the spin valve structures\cite{STTrev1,STTrev2},  but $\mathbf{S}_{i}$ and $\mathbf{S}_{j}$ here represent the spin vectors of two magnetic atoms,  rather than two ferromagnetic layers.   

The coefficients $\beta_{i\leftarrow j}$ and $\gamma_{i\leftarrow j}$ have not been specified yet.  Considering that the non-equilibrium conduction electrons are induced by the electric current,  it is reasonable to assume that $\bm{\sigma}_{i\leftarrow j}$ is proportional to the microscopic current flow $I_{ij}$ between sites $i$ and $j$, namely,  $\beta_{i\leftarrow j}=\lambda_{\beta}^{i\leftarrow j}I_{ij}$,  $\gamma_{j\leftarrow i}=\lambda_{\gamma}^{i\leftarrow j}I_{ij}$  Furthermore,  $\bm{\sigma}_{j\leftarrow i}$ is not necessarily equal to $\bm{\sigma}_{i\leftarrow j}$ because of the unidirectional flow of the electrons.  Therefore,  we introduce the coefficients $\lambda_{\beta}^{\pm}=(\lambda_{\beta}^{i\leftarrow j}\pm\lambda_{\beta}^{j\leftarrow i})/2$,  $\lambda_{\gamma}^{\pm}=(\lambda_{\gamma}^{i\leftarrow j}\pm\lambda_{\gamma}^{j\leftarrow i})/2$ to further decompose the torque $\bm{\tau}_{i\leftarrow j}$ into reciprocal and nonreciprocal terms\cite{SMater}.  Without quantitative details of the coefficients $\lambda_{\beta}^{\pm}$ and $\lambda_{\gamma}^{\pm}$,  the influence of electric current on the spin wave dynamics in the atom scale can already be studied in this general theoretical framework.    

\begin{figure}[!ht]
  \centering
  \includegraphics[width=0.5\textwidth,clip]{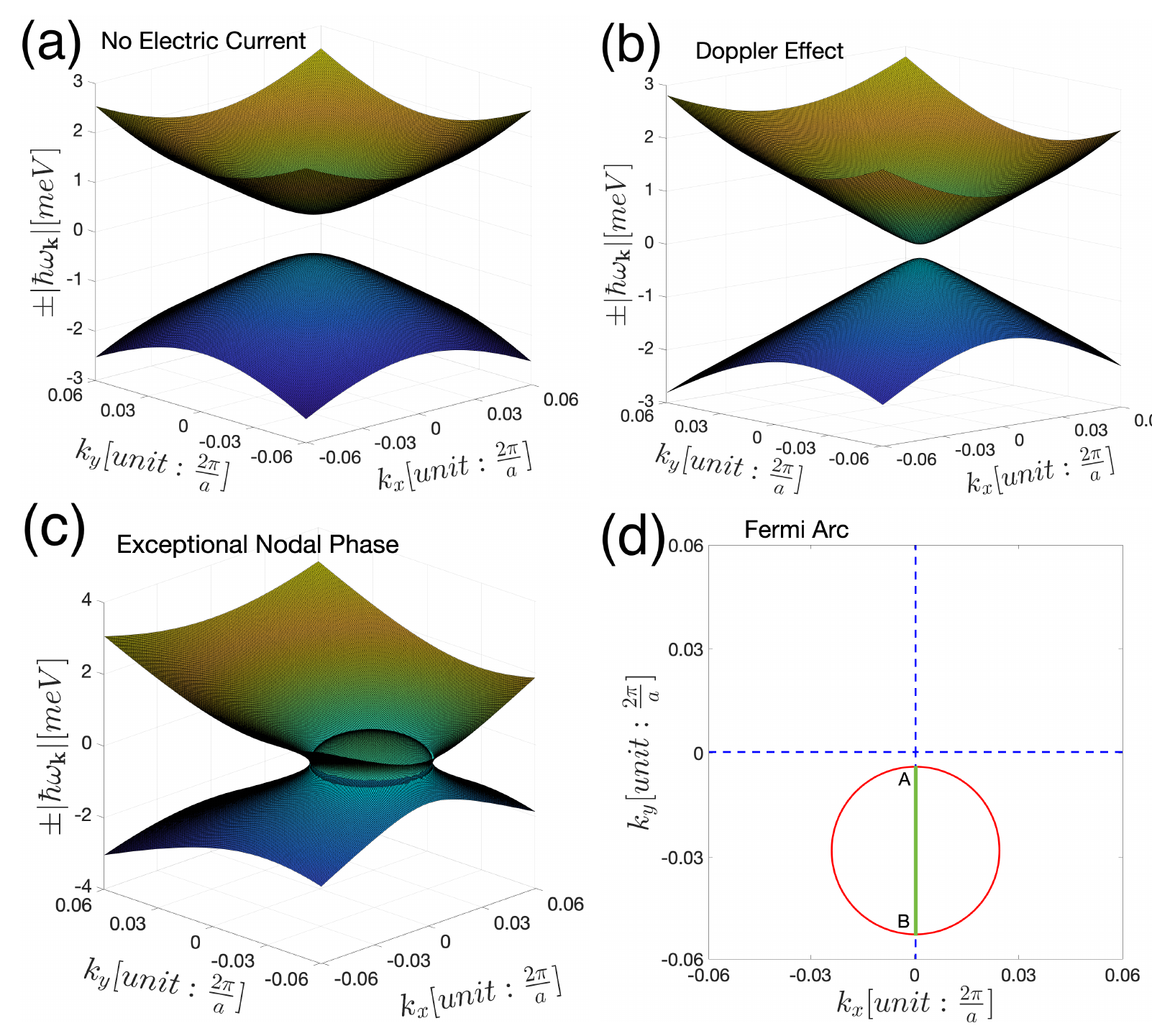}
  \caption{(Color online) Effects of electric current on the spin wave modes of the AFM honeycomb lattice.   Here,  we set $a=6.077$~\AA, $JS=3.85$~meV,  $KS=0.0043$~meV.  (a) Spin wave spectrum without electric current.  (b) Spin wave Doppler shift induced by the $\lambda_{\beta}^{+}$ term.  Here,  $\mathbf{I}_{e}$ is along $y$-direction,  and $I_{e}J_{sd}\lambda_{\beta}^{+}=0.2$~meV.  (c) Exceptional nodal phase induced by $\lambda_{\beta}^{+}$ and  $\lambda_{\beta}^{-}$ terms together.  Here,  $\mathbf{I}_{e}$ is along $y$-direction,  and $I_{e}J_{sd}\lambda_{\beta}^{+}=0.4$~meV,  $I_{e}J_{sd}\lambda_{\beta}^{-}=0.04$~meV. (d) Exceptional points $A$,  $B$ and bulk Fermi arc $\overline{AB}$ (green line) in the nodal phase in case (c).  On the Fermi arc,  $\hbar\omega_{\mathbf{k},+}$ and $\hbar\omega_{\mathbf{k},-}$ will have the same real part,  but different imaginary part. } 
 \label{Fig2}
\end{figure}

\emph{Non-Hermitian Spin Wave Dynamics} The spin wave dynamics of the system is obtained by considering an infinitesimal deviation $\{\mathbf{s}_{i}\}$ from the AFM configuration,  where the localized spins in the sublattice $A$($B$) are aligned along $z$($-z$) direction.   Keeping to the first order of $\{\mathbf{s}_{i}\}$,  we can get the following dynamical equations\cite{SMater}
\begin{eqnarray}
\frac{d}{dt}s_{\mu,A}^{+}&=&-\frac{i}{\hbar}(3J+2K)Ss_{\mu,A}^{+}-\frac{i}{\hbar}JS\sum_{\nu}s_{\nu,B}^{+}\nonumber\\
&&+\frac{i}{\hbar}\sum_{\nu}J_{sd}(-i\beta_{\mu,A\leftarrow\nu,B}+\gamma_{\mu,A\leftarrow\nu,B})s_{\nu,B}^{+},\label{dsadt}\\
\frac{d}{dt}s_{\nu,B}^{+}&=&\frac{i}{\hbar}(3J+2K)Ss_{\mu,B}^{+}+\frac{i}{\hbar}JS\sum_{\mu}s_{\mu,A}^{+}\nonumber\\
&&-\frac{i}{\hbar}\sum_{\mu}J_{sd}(i\beta_{\nu,B\leftarrow\mu,A}+\gamma_{\nu,B\leftarrow\mu,A})s_{\mu,A}^{+}.\label{dsbdt}
\end{eqnarray}
For clarification,  we have used the subscript $\mu,A$ ($\nu,B$) to label the site $A$ ($B$) in the unit cell $\mu$ ($\nu$).   

Substituting the plane wave solution  $s_{\mu,A}^{+}=\sqrt{2S}a_{\mathbf{k}}e^{i\mathbf{R}_{\mu}\cdot\mathbf{k}-i\omega_{\mathbf{k}}t}$ and $s_{\nu,B}^{+}=\sqrt{2S}b_{\mathbf{k}}e^{i\mathbf{R}_{\nu}\cdot\mathbf{k}-i\omega_{\mathbf{k}}t}$ into Eqs.  (\ref{dsadt}) and (\ref{dsbdt}),  we will get the eigen equation
$H_{\mathbf{k}}\mathbf{e}_{\mathbf{k}}=\hbar\omega_{\mathbf{k}}\mathbf{e}_{\mathbf{k}}$,
with the dynamical matrix\cite{SMater}
\begin{eqnarray}
H_{\mathbf{k}}=\left(\begin{array}{cc} (3J+2K)S & JS\gamma_{\mathbf{k}}-I_{e}J_{sd}\lambda_{AB}\eta_{\mathbf{k}} \\
-JS\gamma_{\mathbf{k}}^{*}+I_{e}J_{sd}\lambda_{BA}\eta_{\mathbf{k}}^{*} & -(3J+2K)S 
\end{array}\right)\nonumber
\end{eqnarray}
and the eigen vector $\mathbf{e}_{\mathbf{k}}=\left(\begin{array}{cc}a_{\mathbf{k}}\\b_{\mathbf{k}} \end{array}\right)$.   Here,  we have $\gamma_{\mathbf{k}}\equiv 1+e^{-i\mathbf{a}_{1}\cdot\mathbf{k}}+e^{-i\mathbf{a}_{2}\cdot\mathbf{k}}$,  $\eta_{\mathbf{k}}\equiv j_{1}+j_{2}e^{-i\mathbf{a}_{2}\cdot\mathbf{k}}+j_{3}e^{-i\mathbf{a}_{1}\cdot\mathbf{k}}$,  $\lambda_{AB}\equiv -i\lambda_{\beta}^{+}-i\lambda_{\beta}^{-}+\lambda_{\gamma}^{+}+\lambda_{\gamma}^{-}$,  $\lambda_{BA}\equiv i\lambda_{\beta}^{+}-i\lambda_{\beta}^{-}+\lambda_{\gamma}^{+}-\lambda_{\gamma}^{-}$,  with the notation $j_{l}\equiv\hat{\mathbf{I}}_{e}\cdot\hat{\mathbf{d}}_{l}$.

As a $2\times 2$ traceless non-Hermitian matrix,  $H_{\mathbf{k}}$ can be written as $H_{\mathbf{k}}=\mathbf{d}_{\mathbf{k}}\cdot\bm{\sigma}$ in terms of the Pauli matrices $\bm{\sigma}$.  Here,  both the real and imaginary part of the coefficient $\mathbf{d}_{\mathbf{k}}=\mathbf{d}_{R,\mathbf{k}}+i\mathbf{d}_{I,\mathbf{k}}$ are modulated by the current as\cite{SMater} $\mathbf{d}_{R,\mathbf{k}}=(-I_{e}J_{sd}\lambda_{\gamma}^{-}\Re({\eta_{\mathbf{k}}})-I_{e}J_{sd}\lambda_{\beta}^{-}\Im(\eta_{\mathbf{k}}),I_{e}J_{sd}\lambda_{\gamma}^{-}\Im(\eta_{\mathbf{k}})-I_{e}J_{sd}\lambda_{\beta}^{-}\Re(\eta_{\mathbf{k}}),(3J+2K)S)$,  $\mathbf{d}_{I,\mathbf{k}}=(-I_{e}J_{sd}\lambda_{\gamma}^{+}\Im(\eta_{\mathbf{k}})+I_{e}J_{sd}\lambda_{\beta}^{+}\Re(\eta_{\mathbf{k}})+JS\Im(\gamma_{\mathbf{k}}),  -I_{e}J_{sd}\lambda_{\gamma}^{+}\Re(\eta_{\mathbf{k}})-I_{e}J_{sd}\lambda_{\beta}^{+}\Im(\eta_{\mathbf{k}})+JS\Re(\gamma_{\mathbf{k}}),0)$.  Then the eigen spectrum of $H_{\mathbf{k}}$ will be 
\begin{eqnarray}
\hbar\omega_{\mathbf{k},\pm}=\pm\sqrt{d_{R,\mathbf{k}}^{2}-d_{I,\mathbf{\mathbf{k}}}^{2}+2i\mathbf{d}_{R,\mathbf{k}}\cdot\mathbf{d}_{I,\mathbf{k}}}.\label{hbaromega}
\end{eqnarray}
When there is no electric current,  Eq.~(\ref{hbaromega}) will reduce to the familiar form $\hbar\omega_{\mathbf{k},\pm}=\pm S\sqrt{(3J+2K)^{2}-J^{2}|\gamma_{\mathbf{k}}|^{2}}$\cite{Nernst1,Nernst2},  which has a finite gap $\Delta_{b}\equiv \sqrt{12JKS^{2}+4K^{2}S^{2}}$ at $\Gamma$ point due to the magnetic anisotropy.  For illustration,  Fig.~\ref{Fig2}(a) plots $\pm|\hbar\omega_{\mathbf{k}}|$ in this case with the parameters $a=6.077$~\AA, $JS=3.85$~meV and $KS=0.0043$~meV,  which are taken for a prototypical material MnPS$_{3}$\cite{Nernst1,JPCM1998}.  When a modest current is turned on,  the spin wave spectrum will become complex in general.  Especially,  exceptional points of $\hbar\omega_{\mathbf{k}}$ can appear in the reciprocal space where the two conditions (i) $d_{R,\mathbf{k}}^{2}-d_{I,\mathbf{\mathbf{k}}}^{2}=0$ and (ii) $\mathbf{d}_{R,\mathbf{k}}\cdot\mathbf{d}_{I,\mathbf{k}}=0$ are satisfied simultaneously\cite{RMP2021}.  In the following,  we demonstrate several typical phenomena in the current-modified spin wave dynamics by tuning the parameters of spin torque in $H_{\mathbf{k}}$.

\emph{Doppler Shift and Exceptional Ring}  We find that the reciprocal term $\lambda_{\beta}^{+}$ in the damping-like torque can cause the Doppler shift of spin waves.  As example,  Fig.~\ref{Fig2}(b) plots the numerical solution of Eq.~(\ref{hbaromega}) for an electric current along $y$-direction and $I_{e}J_{sd}\lambda_{\beta}^{+}=0.2$~meV.  Compared with Fig.~\ref{Fig2}(a),  the gap in the spin wave spectrum has become smaller and its location has been shifted.  For an electric current smaller than a critical value $I_{c}=\frac{\sqrt{2}}{3}\frac{\Delta_{b}}{J_{sd}\lambda_{\beta}^{+}}$,   the long wavelength approximation of Eq.~(\ref{hbaromega}) gives the gap value as $\Delta_{b}=\sqrt{\Delta_{b}^{2}-\frac{9}{2}(I_{e}J_{sd}\lambda_{\beta}^{+})^{2}}$ and the gap location as $\mathbf{k}_{g}=-\frac{\sqrt{3}J_{sd}\lambda_{\beta}^{+}}{JSa}\mathbf{I}_{e}$\cite{SMater}.   This current-induced spin wave Doppler shift has been predicted in the spin torque theories for slowly varying magnetization textures\cite{PRB1998,MacDonald2004,AFMtorq2},  and then has been observed in ferromagnetic materials\cite{Science2008,PRB2010,PRL2012}.   Nevertheless,  our results here are not limited to the long wavelength modes around $\Gamma$ point,  but include all the spin wave modes in the whole Brillouin zone.  Such a feature is essential to reveal the current-induced topological phenomena in the spin wave spectrum.
 
When the electric current is larger than the critical value $I_{c}$,   the gap value $\Delta_{b}$ will become imaginary,  and an exceptional ring will appear in the spin wave spectrum.  In this case,  the condition $\mathbf{d}_{R,\mathbf{k}}\cdot\mathbf{d}_{I,\mathbf{k}}=0$ is trivially satisfied in the whole Brillouin zone,  because the dynamical matrix $H_{\mathbf{k}}$ now is pseudo-Hermitian,  namely,  $H_{\mathbf{k}}=\sigma_{z}H_{\mathbf{k}}^{\dag}\sigma_{z}$\cite{RMP2021}.  Then the condition $d_{R,\mathbf{k}}^{2}-d_{I,\mathbf{\mathbf{k}}}^{2}=0$ will result in an exceptional ring centered at $\mathbf{k}_{g}$ with the radius $\frac{1}{JSa}\sqrt{3(I_{e}J_{sd}\lambda_{\beta}^{+})^{2}-\frac{2}{3}\Delta_{b}^{2}}$\cite{SMater}.   As an non-Hermitian topological object,  exceptional rings have been spawned out of Dirac cones by introducing non-Hermitian perturbations in the photonic crystal slabs\cite{Nature2015}.  While in magnetic materials,  we show that exceptional ring in the spin wave spectrum is intrinsically related to the spin wave Doppler effect,  both of which can be induced by tuning the reciprocal term in the damping-like torque.   

\emph{Exceptional Points and Bulk Fermi Arc} The existence of exceptional ring is topologically protected by the pesudo-Hermiticity of $H_{\mathbf{k}}$\cite{RMP2021}.  This constraint will be lifted by turning on the nonreciprocal terms $\lambda_{\beta}^{-}$ or $\lambda_{\gamma}^{-}$ in the torque.  Fig.~2(c) plots the spectrum $\pm|\hbar\omega_{\mathbf{k}}|$ by setting $I_{e}J_{sd}\lambda_{\beta}^{+}=0.4$~meV and $I_{e}J_{sd}\lambda_{\beta}^{-}=0.04$~meV for the electric current along $y$-direction.  In this case,  the solutions of conditions (i) and (ii) form a ring and two crossing lines respectively (see Fig.~\ref{Fig2}(d)),  and their intersections determine two exceptional points $A$ and $B$.  Moreover,  the two branches $\hbar\omega_{\mathbf{k},\pm}$ of the spectrum on the line terminated by $A$ and $B$ have the same real part but different imaginary part,  which forms the ``bulk Fermi arc" in the non-Hermitian topological band theory\cite{ShenZhenFu2018,Science2018}.  In addition,  the exceptional points and the Fermi arc in $\hbar\omega_{\mathbf{k}}$ can be further tuned by the direction of  electric current,  as well as other terms in the spin torque\cite{SMater}.  Therefore,  the electric current can induce rich exceptional nodal phases in the spin wave dynamics in magnetic materials.

\begin{figure}[!ht]
  \centering
  \includegraphics[width=0.5\textwidth,clip]{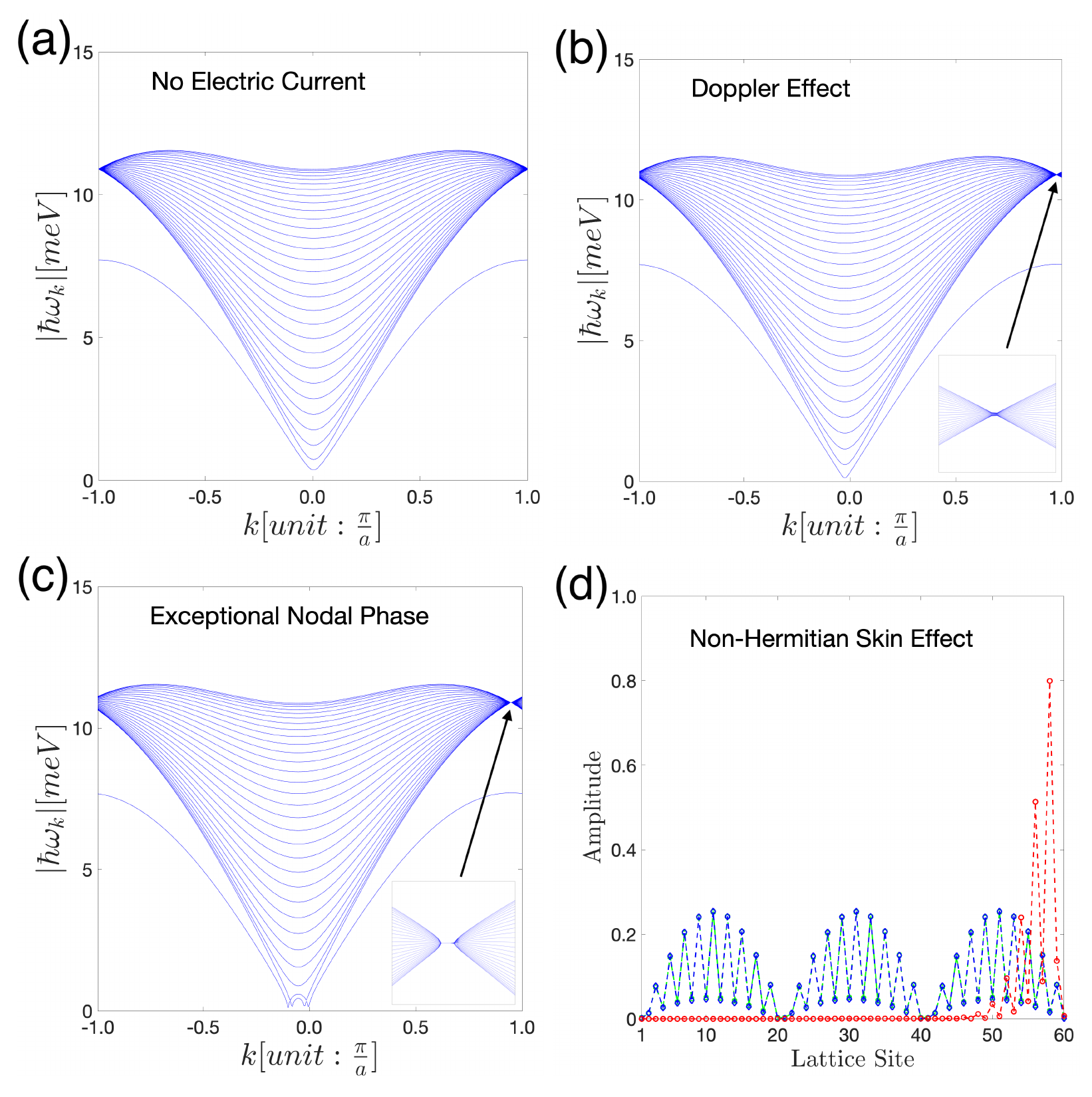}
  \caption{(Color online) Effects of electric current on the spin wave modes of an AFM zigzag nanoribbon.  Here,  we set $N=30$, $a=6.077$~\AA, $JS=3.85$~meV,  $KS=0.0043$~meV.   (a) Spin wave spectrum without electric current.  (b) Spin wave Doppler shift induced by the $\lambda_{\beta}^{+}$ term.  Here,  $I_{e}J_{sd}\lambda_{\beta}^{+}=0.2$~meV.  Inset: $|\hbar\omega_{\mathbf{k}}|$ around the degeneracy point,  which located at $\frac{\pi}{a}$ in case (a) and is now shifted to $\frac{\pi}{a}-\frac{\sqrt{3}J_{sd}\lambda_{\beta}^{+}}{JSa}I_{e}$ by the current.  (c) Higher-order exceptional points and Fermi arc induced by the $\lambda_{\beta}^{-}$ term.  Here,  $I_{e}J_{sd}\lambda_{\beta}^{+}=0.4$~meV and $I_{e}J_{sd}\lambda_{\beta}^{-}=0.04$~meV.   Inset: $|\hbar\omega_{\mathbf{k}}|$ around two high-order exceptional points. (d) Non-Hermitian skin effect induced by the $\lambda_{\beta}^{-}$ term.  Here,  the amplitudes of $\mathbf{s}_{k}$ are plotted for a chosen spin wave mode at $k=\frac{19}{20}\frac{\pi}{a}$ corresponding to the three circumstances: green square for (a); blue diamond for (b); red circle for (c).  } 
 \label{Fig3}
\end{figure}

\emph{Higher-order exceptional points and non-Hermitian skin effect} We also found interesting manifestation of the current-modified spin wave dynamics at the open edges of the AFM lattice.  As example,  nanoribbons with zigzag edges are considered (see Fig.~\ref{Fig1}(b)),  where the lattice sites in the $l$th unit cell are labelled as $\{B_{l,1},A_{l,1},\cdots,B_{l,j},A_{l,j},\cdots,B_{l,N},A_{l,N}\}$.  The plane wave solution of the mode $(k,\omega_{k})$ can be expressed as $s_{l,j,A}^{+}=\sqrt{2S}a_{j,k}e^{ilak-i\omega_{k}t}$ and $s_{l,j,B}^{+}=\sqrt{2S}b_{j,k}e^{ilak-i\omega_{k}t}$,  which will give the eigen equation $\mathcal{H}_{k}\mathbf{s}_{k}=\hbar\omega_{k}\mathbf{s}_{k}$.  Here,  $\mathbf{s}_{k}^{T}=(b_{1,k},a_{1,k},\cdots,b_{j,k},a_{j,k},\cdots,b_{N,k},a_{N,k})$,  and $\mathcal{H}_{k}$ is a $2N\times 2N$ non-Hermitian matrix with the nonzero elements $\mathcal{H}_{k}^{b,j;b,j}=-(3J+2K)S$,  $\mathcal{H}_{k}^{a,j;a,j}=(3J+2K)S$,  $\mathcal{H}_{k}^{b,j;a,j}=-JS(1+e^{-iak})+iJ_{sd}(\lambda_{\beta}^{+}-\lambda_{\beta}^{-})(j_{3}+j_{2}e^{-iak})$,  $\mathcal{H}_{k}^{a,j;b,j}=JS(1+e^{iak})+iJ_{sd}(\lambda_{\beta}^{+}+\lambda_{\beta}^{-})(j_{3}+j_{2}e^{iak})$,  $\mathcal{H}_{k}^{b,j;a,j-1}=-JS$,  $\mathcal{H}_{k}^{a,j;b,j+1}=JS$\cite{SMater}.  Besides,  the open boundary condition is set as $\mathcal{H}_{k}^{b,1;a,0}=\mathcal{H}_{k}^{a,N;b,N+1}=0$ and $\mathcal{H}_{k}^{a,N;a,N}=-\mathcal{H}_{k}^{b,1;b,1}=2J+2K$,  where the coordination number of the edge sites will be $2$ instead of $3$\cite{SMater}.

Fig.~\ref{Fig3}(a) plot the spin wave spectrum $|\hbar\omega_{k}|$ for a $N=30$ nanoribbon with the same parameters in Fig.~\ref{Fig2}(a),  where $\mathbf{I}_{e}=0$.  Because of the reduced coordination number of edge sites,  the dispersion of edge modes is isolated from the spectrum of bulk modes,  and its gap $\Delta_{e}=\sqrt{8JKS^{2}+4K^{2}S^{2}}$ is smaller than the bulk value $\Delta_{b}$\cite{SMater}.  When the electric current is applied along $y$-direction,  the current-induced spin wave Doppler shift will also arise in the nanoribbon structures after turning on the reciprocal term $\lambda_{\beta}^{+}$ in the damping-like torque,  as shown in Fig.~\ref{Fig3}(b).  Here,  the gap is shifted to $-\frac{\sqrt{3}J_{sd}\lambda_{\beta}^{+}}{JSa}I_{e}$ with a smaller value $\sqrt{\Delta_{e}^{2}-3(I_{e}J_{sd}\lambda_{\beta}^{+})^{2}}$\cite{SMater}.  When the electric current is larger than the critical value $\frac{\Delta_{e}}{\sqrt{3}J_{sd}\lambda_{\beta}^{+}}$,  the gap will be closed and exceptional points will appear in the spin wave spectrum\cite{SMater}.  Besides,  there is a degenerate point in the bulk part of the spectrum,  which will also be shifted from $\frac{\pi}{a}$ to $\frac{\pi}{a}-\frac{\sqrt{3}J_{sd}\lambda_{\beta}^{+}}{JSa}I_{e}$ by the electric current\cite{SMater}.  

The nonreciprocal terms in spin torque will also break the pseudo-Hermiticity of $\mathcal{H}_{k}$ for the nanoribbon structures.  After turning on the $\lambda_{\beta}^{-}$ term,  we find that the single degenerate point in the bulk spectrum will become a Fermi arc terminated by two higher-order exceptional points (see inset in Fig.~\ref{Fig3}(c)).  Such a behavior can be mathematically explained by the transfer matrix method\cite{TMatrix},  which gives the bulk spectrum in Fig.~\ref{Fig3}(c) by the expression $\hbar\omega_{k,\pm,\phi}=\pm\sqrt{\Delta_{b}^{2}+8J^{2}S^{2}-\Gamma_{k,+}\Gamma_{k,-}-2JS\sqrt{\Gamma_{k,+}\Gamma_{k,-}}\cos\phi}$ for $\phi_{i=1,\cdots,N-1}\in(0,\pi)$\cite{SMater}.  Here,  we have $\Gamma_{k,+}\Gamma_{k,-}=4\Omega^{2}\cos^{2}(\varphi+\frac{ak}{2})-3(J_{sd}\lambda_{\beta}^{-}I_{e})^{2}\sin^{2}\frac{ak}{2}$ with the notation $\Omega e^{-i\varphi}\equiv JS-i\frac{\sqrt{3}}{2}I_{e}J_{sd}\lambda_{\beta}^{+}$.  If $\lambda_{\beta}^{-}=0$,  we will get $\Gamma_{k,+}\Gamma_{k,-}\geq 0$,  and there is only one degenerate point which satisfies the condition $\cos(\varphi+\frac{ak}{2})=0$; otherwise,  there will be two exceptional points given by the condition $\cos(\varphi+\frac{ak}{2})=\pm\frac{\sqrt{3}J_{sd}\lambda_{\beta}^{-}I_{e}}{2\Omega}\sin\frac{ak}{2}$.  For the $k$ space between these two exceptional points,   $\hbar\omega_{k,\pm,\phi}$ will have the same real part but different imaginary part because of $\Gamma_{k,+}\Gamma_{k,-}<0$,  which interprets the formation of Fermi arc\cite{SMater}.

Non-Hermitian skin effect is another consequence induced by the nonreciprocal terms in the spin torque.  We have examined the spin configurations of the spin wave modes for the three sets of simulation parameters in Fig.~\ref{Fig3}(a-c)\cite{SMater}.  In the cases without the $\lambda_{\beta}^{-}$ term,  the spin wave amplitude will extend over the nanoribbon;  however,   if $\lambda_{\beta}^{-}\neq 0$,   the spin wave amplitude will become localized near one of the nanoribbon edges (see Fig.~\ref{Fig3}(d)).  This anomalous bulk-boundary correspondence\cite{Skin1,Skin2,Skin3,Skin4,Skin5,Skin6,Skin7} directly manifests the non-Hermiticity of the current-modified spin wave dynamics,  which could be utilized to detect the nonreciprocity of the spin torque.

\emph{Conclusion} In a general framework of vector analysis,  we show that an electric current can induce Doppler shift and various non-Hermitian topological phenomena in the spin wave dynamics in a AFM honeycomb lattice.  In turn,  the reciprocal and nonreciprocal terms in the spin torque could be revealed by measuring these effects of electric current on the spin wave dynamics.  Furthermore,  we anticipate that similar phenomena could widely exist in other magnetic materials with different lattice structure,  magnetic order,  or higher dimension,  which therefore serve as potential platforms to explore fundamental physics and develop frontier technology of non-Hermitian magnons.    

The work is supported by the National Natural Science Foundation of China (Grants No. 12074195),  the Croucher Senior Research Fellowship,  and the Research Grant Council of Hong Kong SAR (HKU17303518).


\begin{thebibliography}{99}
\bibitem{Callaway} J.  Callaway,  \emph{Quantum Theory of the Solid State}(Academic Press,  1991).
\bibitem{JMMM1999} J.C.  Slonczewski,  J.  Magn.  Magn.  Mater.  \textbf{195},  L261 (1999). 
\bibitem{PRL1998} M. Tsoi,  A. G. M.  Jansen,  J. Bass,  W.-C.  Chiang,  M. Seck,  V. Tsoi,  and P. Wyder,  Phys. Rev. Lett.  \textbf{80}, 4281 (1998).
\bibitem{Nature2000} M. Tsoi,  A. G. M.  Jansen,  J. Bass,  W.-C.  Chiang,  V. Tsoi,  and P. Wyder,  Nature,  \textbf{406},  46 (2000).
\bibitem{PRL2003} Y.  Ji,  C. L.  Chien,  and M. D. Stiles,  Phys.  Rev.  Lett.  \textbf{90},  106601 (2003).
\bibitem{PRL2012Xiao} J.  Xiao and G. E.W.  Bauer,  Phys. Rev.  Lett. \textbf{108},  217204 (2012).
\bibitem{NC2018Akerman}  A.  Houshang,  R. Khymyn,  H. Fulara,  A. Gangwar,  M. Haidar,  S.R. Etesami,  R. Ferreira,  P.P. Freitas,  M. Dvornik,  R.K. Dumas,  and J.  \AA kerman,  Nat.  Commun.  \textbf{9},  4374 (2018).
\bibitem{PRB1998} Y. B. Bazaliy,  B.A. Jones,  and S.-C. Zhang,  Phys. Rev. B \textbf{57}, R3213 (1998).
\bibitem{MacDonald2004} J. Fern\'{a}ndez-Rossier,  M.  Braun,  A. S.  N\'{u}\~nez,  and A. H. MacDonald,  Phys.  Rev.  B \textbf{69},  174412 (2004).
\bibitem{Science2008} V. Vlaminck and M. Bailleul,  Science \textbf{322}, 410 (2008).
\bibitem{PRB2010} M. Zhu,  C. L.  Dennis,  and R. D.  McMichael,  Phys. Rev.  B \textbf{81}, 140407(R) (2010).  
\bibitem{PRL2012} K.  Sekiguchi,  K.  Yamada,  S.-M.  Seo,  K.-J.  Lee,  D.  Chiba,  K.  Kobayashi,  and T.  Ono,  Phys.  Rev.  Lett.  \textbf{108},  017203 (2012).
\bibitem{PRL2009} S.-M.  Seo,  K.-J.  Lee,  H.  Yang,  and T.  Ono,  Phys.  Rev.  Lett.  \textbf{102}, 147202 (2009).
\bibitem{PRL2014} A. Hamadeh, O.  d'Allivy Kelly,  C.  Hahn,  H.  Meley,  R. Bernard,  A. H.  Molpeceres,  V. V.  Naletov,  M.  Viret,  A. Anane,  V. Cros,  S. O. Demokritov,  J.  L. Prieto,  M. Mu\~noz,  G.  de Loubens,  and O.  Klein,  Phys.  Rev.  Lett.  \textbf{113}, 197203 (2014).
\bibitem{JAP2020} A.  Mahmoud,  F.  Ciubotaru,  F.  Vanderveken,  A. V.  Chumak,  S.  Hamdioui,  C.  Adelmann,  and S.  Cotofana,  J.  Appl.  Phys.  \textbf{128}, 161101 (2020).
\bibitem{NC2014} A. V.  Chumak,  A.  A.   Serga,  B.  Hillebrands,  Nat.  Commun.  \textbf{5},  4700 (2014).
\bibitem{NatPhys2015} A. V. Chumak,  V. I.  Vasyuchka,  A. A.  Serga,  and B. Hillebrands,  Nature Physics,  \textbf{11},  453(2015). 
\bibitem{Nature1}  C.  Gong,  L.  Li,  Z.  Li,  H.  Ji,  A.  Stern,  Y.  Xia,  T.  Cao,  W.  Bao,  C.  Wang,  Y.  Wang,  Z. Q. Qiu,  R. J.  Cava,  S. G.  Louie,  J.  Xia,  and  X.  Zhang,  Nature,  \textbf{546},  265 (2017). 
\bibitem{Nature2} B.  Huang,  G.  Clark,  E.  Navarro-Moratalla,  D. R.  Klein,  R.  Cheng,  K. L.  Seyler,  D.  Zhong,  E.  Schmidgall,  M. A.  McGuire,  D. H.  Cobden,  W.  Yao,  D.  Xiao,  P.  Jarillo-Herrero,  and X.  Xu,  Nature,  \textbf{546},  270(2017).
\bibitem{NC2018} W.  Jin,  H. H.  Kim,  Z.  Ye,  S.  Li,  P.  Rezaie,  F.  Diaz,  S.  Siddiq,  E.  Wauer,  B.  Yang,  C.  Li,  S.  Tian,  K.  Sun,  H.  Lei,  A. W.  Tsen,  L.  Zhao,   and R.  He,  Nat.  Commun.  \textbf{9} 5122 (2018).  
\bibitem{PRX2018} L.  Chen,  J.-H.  Chung,  B.  Gao,  T.  Chen,  M. B.  Stone,  A. I.  Kolesnikov,  Q.  Huang,  and P.   Dai,  Phys.  Rev.  X \textbf{8}, 041028 (2018).
\bibitem{PRX2019} W.  Xing,  L.  Qiu,  X.  Wang,  Y.  Yao,  Y.  Ma,  R.  Cai,  S.  Jia,  X. C.  Xie,  and W.  Han,  Phys.  Rev.  X \textbf{9}, 011026 (2019).
\bibitem{Nernst1} R.  Cheng,  S.  Okamoto,  and D.  Xiao,  Phys.  Rev.  Lett.  \textbf{117}, 217202 (2016).
\bibitem{Nernst2} V. A.  Zyuzin and A. A.  Kovalev,  Phys.  Rev.  Lett.  \textbf{117}, 217203 (2016).
\bibitem{NatMat2020} X.-X.  Zhang,  L.  Li,  D.  Weber,  J.  Goldberger,  K. F.  Mak,   and J.  Shan,  Nat.  Mater.  \textbf{19},  838 (2020).
\bibitem{NC2021} G.  Chen,  S.  Qi,  J.  Liu,  D.  Chen,  J.  Wang,  S.  Yan,  Y.  Zhang,  S.  Cao,  M.  Lu,  S.  Tian,  K.  Chen,  P.  Yu,  Z.  Liu,  X. C.  Xie,  J.  Xiao,  R.  Shindou,  and  J.-H.  Chen,  Nat.  Commun.  \textbf{12},  6279(2021).
\bibitem{SA2021} F.  Zhu,  L.  Zhang,  X.  Wang,  F.  Jos dos Santos,  J.  Song,  T.  Mueller,  K.  Schmalzl,  W. F.  Schmidt,  A.  Ivanov,  J. T.  Park, J.  Xu,  J.  Ma,  S.  Lounis,  S.  Bl\"{u}gel,   Y.  Mokrousov,  Y. Su,  and T.  Br\"{u}ckel,  Sci.  Adv.  \textbf{7},  eabi7532 (2021).
\bibitem{STT1} J. C.  Slonczewski,  J.  Magn.  Magn.  Mater. \textbf{159},  L1 (1996).
\bibitem{STT2} L.  Berger,  Phys. Rev. B \textbf{54}, 9353 (1996).
\bibitem{STTrev1} D.C.  Ralpha and M.D.Stiles,  J.   Magn.   Magn.  Mater.  \textbf{320},  1190 (2008).
\bibitem{STTrev2} A.  Brataas,  A. D. Kent,  and H.  Ohno,  Nat.  Mater.  \textbf{11},  372 (2012). 
\bibitem{ZhangLi2004} S. Zhang and Z. Li,  Phys.  Rev.  Lett.  \textbf{93},  127204(2004).
\bibitem{AFMtorq1} K. M. D. Hals, Y.  Tserkovnyak,  and A.  Brataas,  Phys.  Rev.  Lett.  \textbf{106}, 107206 (2011).
\bibitem{AFMtorq2} A. C.  Swaving and R. A.  Duine,  Phys.  Rev.  B \textbf{83}, 054428 (2011).
\bibitem{AFMtorq3} E. G.  Tveten,  A.  Qaiumzadeh,  O. A. Tretiakov,  and A.  Brataas,  Phys.  Rev.  Lett.  \textbf{110}, 127208 (2013).
\bibitem{AFMtorq4}  R.  Cheng and Q. Niu,  Phys.  Rev.  B \textbf{89}, 081105(R) (2014). 
\bibitem{AFMtorq5} R.  Cheng,  J.  Xiao,  Q.  Niu,  and A.  Brataas,  Phys.  Rev.  Lett.  \textbf{113},  057601 (2014).
\bibitem{AFMtorq6} Y.  Yamane,  J.  Ieda,  and J.  Sinova,  Phys.  Rev.  B \textbf{94}, 054409 (2016).
\bibitem{AFMtorq7} J.  Fujimoto,  Phys.  Rev.  B \textbf{103}, 014436 (2021).
\bibitem{Mahan} G.D. Mahan,  \emph{Many-Particle Physics}(Kluwer Academic/Plenum Publishers,  New York,2000). 
\bibitem{SMater} See Supplemental Material at [URL will be inserted by publisher] for more theoretical details and numerical examples. 
\bibitem{JPCM1998} A.  R.  Wilde,  B. Roesslix,  B.  Lebechk,  and K. W.  Godfreyy,  J.  Phys.  Condens.  Matter \textbf{10},  6417 (1998).
\bibitem{RMP2021} E. J.  Bergholtz,  J.  C.  Budich,  and F.  K.  Kunst,  Rev.  Mod.  Phys.  \textbf{93},  015005 (2021).
\bibitem{Nature2015} B.  Zhen,  C. W.  Hsu,  Y.  Igarashi,  L.  Lu,  I.  Kaminer,  A.  Pick,  S.-L.  Chua, J. D. Joannopoulos,  and M.  Solj\u{a}ci\u{c},  Nature,  \textbf{525},  354 (2015).
\bibitem{ShenZhenFu2018} H.  Shen,  B.  Zhen,  and L.  Fu,   Phys.  Rev.  Lett.  \textbf{120},  146402 (2018).
\bibitem{Science2018} H.  Zhou,  C.  Peng,  Y.  Yoon,  C. W.  Hsu,  K. A. Nelson,  L. Fu,  J. D.  Joannopoulos,  M.  Solja\u{c}i\u{c},  and B.  Zhen,  Science,  \textbf{359},  1009 (2018).
\bibitem{TMatrix} F. K.  Kunst and V.  Dwivedi,  Phys.  Rev.  B \textbf{99}, 245116 (2019).  
\bibitem{Skin1} S. Yao and Z.  Wang,  Phys. Rev.  Lett. \textbf{121}, 086803 (2018). 
\bibitem{Skin2} S.  Yao,  F.  Song,  and Z.  Wang,  Phys.  Rev.  Lett. \textbf{121},  136802 (2018).
\bibitem{Skin3} F. K.  Kunst,  E.  Edvardsson,  J.C. Budich,  and  E.J.  Bergholtz,  Phys.  Rev.  Lett.  \textbf{121},  026808 (2018).
\bibitem{Skin4} L.  Jin and Z.  Song,  Phys.  Rev.  B \textbf{99},  081103(R) (2019).
\bibitem{Skin5} H. Q.  Wang,  J. W. Ruan,  and H. J.  Zhang,  Phys.  Rev.  B \textbf{99},  075130 (2019).
\bibitem{Skin6} Z. Yang,  K.  Zhang,  C.  Fang,  and J. P.  Hu,  Phys.  Rev.  Lett.  \textbf{125},  226402 (2020).
\bibitem{Skin7} Z.  Zhang, Z.  Yang,  and J. P.  Hu,  Phys.  Rev.  B \textbf{102},  045412 (2020).


\end{thebibliography}
\end{document}